\documentclass[a4paper, 11pt]{article}

\usepackage {amssymb,latexsym,amsmath, amsthm, graphicx, amsfonts, verbatim}

\usepackage{subcaption}
\RequirePackage[OT1]{fontenc}
\RequirePackage{amsthm,amsmath}
\RequirePackage[numbers]{natbib}
\RequirePackage[colorlinks,citecolor=blue,urlcolor=blue]{hyperref}

\numberwithin{equation}{section}
\theoremstyle{plain}
\newtheorem{thm}{Theorem}[section]

\title{Regression Trees for Longitudinal Data}

\author{Madan Gopal Kundu and Jaroslaw Harezlak\\
\\
Indiana University Fairbank School of Public Health, Indianapolis, IN}
\date{\today}

\begin{document}
\maketitle

\begin{abstract}
 While studying response trajectory, often the population of interest may be diverse enough to exist distinct subgroups within it and the longitudinal change in response may not be uniform in these subgroups. That is,  the timeslope and/or  influence of covariates in longitudinal profile may vary among these different subgroups. For example, \citet{raudenbush2001comparing} used depression as an example to argue that it is incorrect to assume that all the people in a given population would be experiencing either increasing or decreasing levels of depression. In such cases, traditional linear mixed effects model  (assuming common parametric form for covariates and time) is not directly applicable for the entire population as a group-averaged trajectory can mask important subgroup differences.  \textcolor{black}{Our aim is to identify and characterize longitudinally homogeneous subgroups based on the combination of baseline covariates in the most parsimonious way. This goal can be achieved via constructing regression tree for longitudinal data using baseline covariates as partitioning variables.  We have proposed LongCART algorithm to construct regression tree for the longitudinal data.  In each node, the proposed LongCART algorithm determines the need for further splitting (i.e.  whether parameter(s) of longitudinal profile is influenced by any baseline attributes) via {\it parameter instability tests} and thus the decision of further splitting is type-I error controlled. We have obtained the asymptotic results for the proposed instability test and examined finite sample behavior of the whole algorithm through simulation studies. Finally, we have applied the LongCART algorithm to study the  longitudinal changes in  {\it choline} level among HIV patients. }
\end{abstract}


\section{Introduction}\label{sec:intro}
In longitudinal studies, repeated measurements of the outcome variable are often collected at irregular and possibly subject-specific time points. Parametric regression methods for analyzing such data have been developed by \citet{laird1982random} and \citet{liang1986longitudinal} among others, and have been summarized by \citet{diggle2002analysis}. If the population under consideration is diverse and there exists several distinct subgroups within it, the true parameter value(s) for longitudinal mixed effects model may vary between these subgroups. In such cases, the traditional mixed effects models, for example linear mixed effects model, which assume a common parametric form for the covariates and time are not  appropriate.  For example,  \citet{raudenbush2001comparing} used a longitudinal depression study as an example to argue that it is incorrect to assume that all the people in a given population will be experiencing either increasing or decreasing levels of depression.  In such instances, an assumption of a common parametric form will mask important subgroup differences and will lead to erroneous conclusions. In our work, our interest is to identify meaningful and interpretable subgroups with differential longitudinal trajectories and/or differential covariate effects' on the response variable from such a heterogeneous population. We propose a regression tree construction technique for the longitudinal data using baseline characteristics as partitioning variables, LongCART algorithm, that (1) takes the decision about further splitting at each node controlling type I error, and (2) is applicable in cases when measurements are taken at subject specific time-points.
\begin{figure}
\centerline{%
\includegraphics [angle=0,width=130mm, height=40mm]{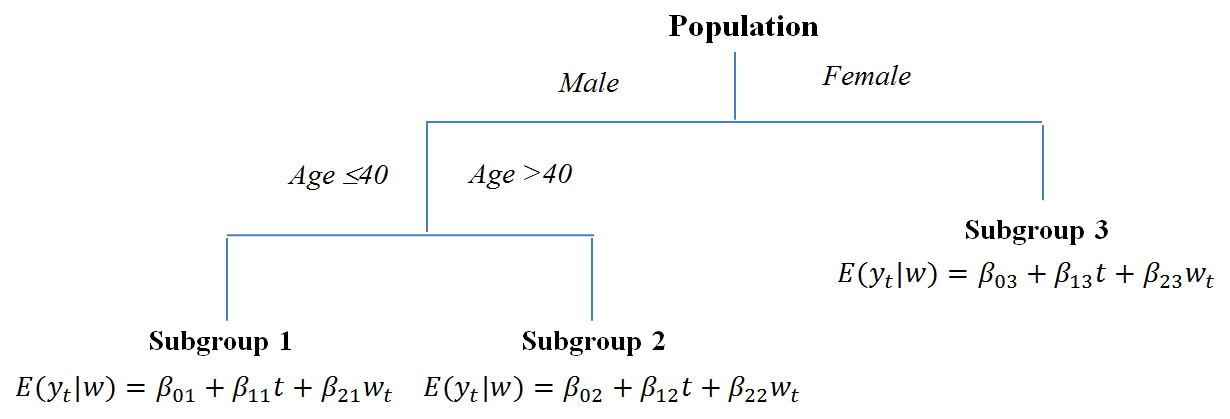}}
\caption{Sample longitudinal tree. The population consists of 3 subgroups and they differ in their longitudinal profiles. These subgroups are defined by the partitioning variables gender and age.}
\label{samptree}
\end{figure}
When the longitudinal profile in a population depends on the baseline covariates, the most common strategy is to include these covariates and their interactions with the time-varying covariate in the model. However, this strategy has some inherent drawbacks such as, over-fitting due to inclusion of all possible interaction terms, requires to specify functional form of the association with baseline covariates, and cannot capture nonlinear effect of baseline covariates. Our goal is to determine the most parsimonious model consisting of a number of homogeneous subgroups from a heterogeneous population profile without strict parametric restrictions or prior information.
One of the popular technique to construct homogeneous subgroups is latent class modeling (LCM) \citep{muthen1999finite}. 
An alternative approach is to construct a regression tree with longitudinal data \citep{segal1992tree}. Advantages of regression tree technique over LCM are: (1) it  characterizes the subgroups in terms of partitioning variables and (2) the number of subgroups need not to be known a-priori. In general, the thrust of any tree technique is the extraction of meaningful subgroups characterized by common covariate values and homogeneous outcome. For longitudinal data, this homogeneity can pertain to the mean and/or covariance structure \citep{segal1992tree}.

Throughout this article, we refer to the regression tree with longitudinal data as `{\it longitudinal  tree}'. Figure~\ref{samptree} displays a  toy example of longitudinal tree. This longitudinal tree represents a heterogeneous population with three distinct subgroups in terms of their longitudinal profiles. These subgroups can be characterized by {\it gender} and {\it age}. Here, {\it gender} and {\it age} are baseline attributes. 
In each of the three subgroups, the longitudinal trajectory depends on the covariates $w_1, \ldots, w_q$, but these subgroups are heterogeneous in terms of the true coefficients associated with their longitudinal profiles.  Consider the following form of a linear longitudinal mixed effects model
\begin{equation}
y_{it}=\beta_0^x + \beta_1^x t+ \mathbf{w}_{it}^{\top} \boldsymbol{\beta}^x +\mathbf{z}_{it}^{\top}\mathbf{b}_i +\epsilon_{it} \label{eq:initial}
\end{equation}
where $i$ is the subject index and $y$, $t$ and $\mathbf{w}$ denote the outcome variable, time and the vector of measurements of scalar covariates $w_1, \ldots, w_q$, respectively.
Let $X_1^{G_1}, \ldots, X_S^{G_S}$ include all potential baseline attributes (with cut-off points $G_1, \cdots, G_S$, respectively) that might influence the longitudinal trajectory in \eqref{eq:initial}. The superscript $x$ is added to the coefficients $\beta_0, \beta_1$ and $\boldsymbol{\beta}$ to reflect their possible dependence on these baseline attributes. Let  ${\boldsymbol{\theta}^x}=(\beta_0^x, \beta_1^x, {\boldsymbol{\beta}^x})^{\top}$. With such a model, `{\it homogeneity}' refers to the situation when the true value of ${\boldsymbol{\theta}^x}$ remains  same for all the individuals in the entire population, {\it i.e.} $\boldsymbol{\theta}^x=\boldsymbol{\theta}$. When the longitudinal changes in the population of interest are {\it heterogeneous} there exists distinct subgroups differing in terms of the true values of the coefficients, {\it i.e.} $\boldsymbol{\theta}^x\neq \boldsymbol{\theta}$. To model influence of $\{X_1^{G_1}, \ldots, X_S^{G_S}\}$ on the longitudinal trajectory of $y$ non-parametrically, we have used these baseline attributes as the partitioning variables for construction of longitudinal  tree. \\

In constructing a longitudinal tree through binary partitioning, one way to choose a partition is via maximizing improvement in goodness of fit criterion. For example, \citet{abdolell2002binary} chose deviance as a goodness of fit criterion. They evaluated deviance at each split of a given partitioning variable and selected the partition with maximum reduction in deviance for the binary splitting.  However, this procedure requires a large number of statistical tests. With $S$ partitioning variables: $X_1^{G_1}, \ldots, X_S^{G_S}$, with cut-off points $G_1, \ldots, G_S$, respectively, the total number of tests is $\sum_{s=1}^S{(G_s-1)}$. Clearly, this approach leads to the multiple testing problem. To minimize the problem of multiplicity, we propose the LongCART algorithm for construction of regression tree that involves only single test for each partitioning variable. We call such a test: {\it ``test for parameter instability"}. Hence, with $S$ partitioning variables, we need to perform only $S$ {\it tests for parameter instability}. The number of tests with the proposed approach is much smaller than $\sum_{s=1}^S{(G_s-1)}$ in presence of continuous partitioning variables and/or categorical partitioning variables with more than two levels. Consequently, LongCART algorithm puts a better check on the type I error. 

\textcolor{black}{We want to construct the regression with a certain level of confidence. In general, controlling type I error rate for the entire tree construction process is very difficult. First, the number of branches is unknown a-priori, and second, there is a large number of possibilities of choosing a split. To address the issue of type I error, at each node, we divide the task of finding the best split at a given node into two sub-tasks: (a) first, we identify if there is a need for further splitting and (b) second, given there is need for further splitting, we choose the optimum splitting point. Our proposed LongCART algorithm controls the type I error while performing the first task. Once this decision is made, the optimum split is chosen without additional testing. In order to get a better intuition for the LongCART algorithm, let us assume there is only a single partitioning variable, say $X^G$, with $G$ cut-off points. In such case, LongCART algorithm identifies the best split at a given node in a two-step process as follows:}
\begin{itemize} 
\item {{\bf Step 1.}}  Perform an overall  {\it parameter instability test} to detect any evidence of heterogeneity of longitudinal model parameters across $G$ cut-off points of $X^G$.
\item {{\bf Step 2.}} Given that there is a `significant' evidence for heterogeneity, the split that provides maximal improvement in goodness of fit criterion is chosen as a cut-off point for the tree construction.
\end{itemize}
\textcolor{black}{We adapt the LongCART algorithm in situations with multiple partitioning variables via repeating the {\it parameter instability test} for each partitioning variable controlling type I error at a given level.
We continue to the second step using the `most significant' partitioning variable. Details of this algorithm are presented in Section~\ref{TreeConstruction}. The key idea here is that we are combining the multiple testing procedure (step 1) with model selection (step 2) in order to control the type I error while taking the decision on splitting at each node.}

In order to construct a test for {\it parameter instability}, we borrow an idea from the time-series literature. In time-series context often the goal is to evaluate whether the parameter of a regression model is stable across different time points. This is often known as a {\it test for structural change} or {\it constancy of parameters} \citep[e.g.,][]{brown1975techniques, nyblom1989testing,  hjort2002tests}. We apply a similar idea to evaluate whether the true values of the parameter  remains the same across the cut-off values of a partitioning variable in a mixed effects longitudinal model of interest.


In this paper we utilize the parameter instability test in multiple ways. First, in the case of continuous partitioning variables, the proposed test uses the  results on score process derived by \citet{hjort2002tests} in conjunction with the properties of Brownian motion and Brownian Bridge. Second, for categorical partitioning variables with a small number of cut-off points, a test for parameter instability is derived  in a straightforward way by employing asymptotic normality of the score functions. We derive the asymptotic properties of the instability test and explore its size and power through an extensive simulation study. Finally, we use these instability tests to construct an algorithm for regression trees with longitudinal data.

Among the tree based methods, classification and regression tree (CART) methods \citep{breiman1984classification} is probably the most popular one. \citet{zeileis2008model} have extended the concept of CART methodology in the context of fitting cross-sectional regression models. Binary partitioning for longitudinal data has been proposed  first by \citet{segal1992tree}. However, Segal's implementation is restricted to longitudinal data with a regular structure, that is all the subjects have an equal number of repeated observations at the same time points \citep{zhang1999recursive}.
\citet{zhang1997multivariate} proposes  multivariate adaptive splines to analyze longitudinal data. Their method, multivariate adaptive splines for the analysis of longitudinal data (MASAL), can be used to generate regression trees for longitudinal data. \citet{abdolell2002binary}  used deviance as a goodness-of-fit criterion for binary partitioning. They controlled the level of Type I error via permutation test taking into account testing multiplicity. However, permutation tests are computer intensive and the time taken to fit the models is intimidatingly high even for medium-sized data. \citet{sela2012re} as well as \citet{galimberti2002regression} merged the subgroup differences with the random individual differences. They constructed the regression tree through an iterative two-step process. In the first step, they obtained the random effects' estimates and in the second step, they constructed the regression tree ignoring the longitudinal structure. They repeat these two steps until the estimates of the random effect converge in the first step. \textcolor{black}{The LongCART algorithm provides an improvement over the existing methods in the following aspects: (1) the test for the decision about further splitting at each node is type I error controlled, (2) it is applicable when the measurements are taken at the subject-specific time points, (3) it does not merge the group differences with the random subject effect components and (4) it reduces computational time.}

The remainder of this paper is organized as follows. In Section \ref{PrelimLongTree} the longitudinal mixed effects models of interest are  summarized. Tests for parameter instability  for continuous and categorical partitioning variable cases are discussed separately in Section \ref{InstabTest}. Algorithm for constructing regression trees along with measures of improvement and a pruning technique are discussed in Section \ref{TreeConstruction}. Results from the simulation studies examining the performance of the instability test and the regression tree as a whole are reported in Section \ref{SimulationRegTree}. An application of the longitudinal regression tree method is illustrated on the metabolite data collected from the chronically HIV-infected patients in Section \ref{TreeApplication}.

\section{Notation and statistical model}\label{PrelimLongTree}
Let $\{y_{it}, \mathbf{w}_{it}\}$ be a set of measurements recorded on the $i^{th}$ subject ($i=1,\dots,N$) at time $t = (t_1,\dots,t_{n_i})$, where $y$ is a continuous scalar outcome; and $\mathbf{w}$ is the vector of measurements on scalar covariates $w_1, \ldots, w_q$. We assume that these covariates are linearly associated with $y$. In addition, for each individual, we observe a vector of attributes $(X^{G_1}_{1i}, \ldots, X^{G_S}_{Si})$ measured at baseline. We assume that $X_1^{G_1}, \ldots, X_S^{G_S}$ includes all potential baseline attributes that can influence  the longitudinal trajectory of $y$ and its association with covariates $w_1, \ldots, w_q$. Further, we do not assume the strict functional form of these baseline attributes' influence. We use the variables  $X_1^{G_1}, \ldots, X_S^{G_S}$ as the candidate partitioning variables to construct a longitudinal regression tree to discover meaningful and interpretable subgroups with differential changes in $y$ characterized by the $X_1^{G_1}, \ldots, X_S^{G_S}$.

When the longitudinal profile is homogeneous in the entire population,  we can fit the following traditional linear mixed effects model for all $N$ individuals \citep{laird1982random}
\begin{equation}\label{eq:longmod}
y_{it}=\beta_0 + \beta_1 t+ \mathbf{w}_{it}^{\top} \boldsymbol{\beta} +\mathbf{z}_{it}^{\top}\mathbf{b}_i +\epsilon_{it},
\end{equation}
where $\epsilon_{it} \sim  N(0,\sigma^2)$ and $\mathbf{b}_i$ is the vector of
random effects pertaining to subject $i$ and distributed as $N(0, \sigma^2\mathbf{D})$.  By `{\it homogeneity}' we mean that the true value of $\mathbf{\boldsymbol{\theta}}^{\top}=(\beta_0, \beta_1, \boldsymbol{\beta}^{\top})$ remains the same for all the individuals in the population. In fact, \eqref{eq:longmod} is the simplified version of model  in \eqref{eq:initial} under homogeneity.

We follow the common assumptions made in longitudinal modeling that $\mathbf{z}_{it}$ is a subset of $[\mathbf{w}_{it}^{\top} \;\; t]^{\top}$; $\epsilon_{it}$ and $\mathbf{b}_i$ are independent;  $\epsilon_{it}$ and $\epsilon_{i't'}$ are independent whenever $i \neq i'$ or $t \neq t'$ or both, and $\mathbf{b}_i$ and $\mathbf{b}_{i'}$ are independent if $i \neq i'$. Here,  $\mathbf{w}_{it}^{\top}\mathbf{\beta}$ is the  fixed effect term and $\mathbf{z}_{it}^{\top}\mathbf{b}_i$ is the standard random effects term. For the $i^{th}$ subject, we rewrite the Eq. \eqref{eq:longmod} as follows
\begin{equation}\label{eq:longmod2}
\mathbf{y}_i=\mathbf{w}_i\mathbf{\boldsymbol{\theta}} +\mathbf{z}_i\mathbf{b}_i +\mathbf{\epsilon}_i,
\end{equation}
where $\mathbf{y}_i^{\top}=(y_{i1}, \ldots, y_{in_i})$, $\mathbf{w}_i$ is the design matrix consisting of the intercept, time ($t$) and covariates ($\mathbf{w}$). $n_i$ is the number of  observations obtained from the $i^{th}$ individual. The score function for estimating $\mathbf{\boldsymbol{\theta}}$ under \eqref{eq:longmod2} is  \citep[see e.g.,][]{demidenko2004mixed}
\begin{equation*}
\mathbf{u}(\mathbf{y}_i, \mathbf{\boldsymbol{\theta}})=\frac{d}{d\mathbf{\boldsymbol{\theta}}}l(\mathbf{y}_i, \mathbf{\boldsymbol{\theta}})=\frac{1}{\sigma^2}\mathbf{w}_i^{\top}\mathbf{V}_i^{-1}(\mathbf{y}_i - \mathbf{w}_i\mathbf{\boldsymbol{\theta}})
\end{equation*}
where $\mathbf{V}_i =\mathbf{I}+\mathbf{z}_i\mathbf{D}\mathbf{z}_i^{\top}$ and $\mathbf{e}_i=\mathbf{y}_i-\mathbf{w}_i \mathbf{\boldsymbol{\theta}}$.  Further, its variance is
\begin{equation*}
\mbox{Var}\left[\mathbf{u}(\mathbf{y}_i, \mathbf{\boldsymbol{\theta}})\right]=\mathbf{J}(\mathbf{\boldsymbol{\theta}})=-E\left[\frac{d}{d\mathbf{\boldsymbol{\theta}}}\mathbf{u}(\mathbf{y}_i, \mathbf{\boldsymbol{\theta}})\right]=\frac{1}{\sigma^2}\mathbf{w}_i^{\top}\mathbf{V}_i^{-1}\mathbf{w}_i
\end{equation*}
%
%
Likelihood estimate of $\mathbf{\boldsymbol{\theta}}$ obtained using all the observation from $N$ subjects is valid only if the entire population under consideration is homogeneous.   If the entire population is not homogeneous in terms of $\mathbf{\boldsymbol{\theta}}$  then the likelihood estimate obtained considering all the subjects together are misleading; the extent and direction of ambiguity in the estimate will depend on the nature and proportion of heterogeneity in the sampled individuals. Therefore, under the assumption that $X_1^{G_1}, \ldots, X_S^{G_S}$ are the only attributes that influences the longitudinal profiles of $y$,  it is important to decide first whether the true value of $\mathbf{\boldsymbol{\theta}}$ remains the same for all the all the subgroups defined by $X_1^{G_1}, \ldots, X_S^{G_S}$ or not. In the next section, we describe  statistical tests to assess whether the true value of $\mathbf{\boldsymbol{\theta}}$ remains the same across all the values of a given partitioning variable.
%
\section{Test for parameter instability}\label{InstabTest}
The purpose of  {\em parameter instability test} is to test whether the true value of $\mathbf{\boldsymbol{\theta}}$ remains the same across all distinct values of baseline attributes (i.e. partitioning variables). Let $X^G\in \{X_1^{G_1}, \ldots, X_S^{G_S}\}$ be any partitioning variable with $G$ ordered cut-off points: $c_{(1)}< \ldots < c_{(G)}$ and $\boldsymbol{\theta}_{(g)}$ be the true value of $\boldsymbol{\theta}$ when $X^G=c_{(g)}$. Assume that there are $m_g$ subject with  $X^G=c_{(g)}$. We denote the cumulative number of subjects with $X^G\leq c_{(g)}$ by $M_g$. That is, $M_g=\sum_{j=1}^g{m_j}$ and  $M_G=\sum_{j=1}^G{m_j}=N$. We want to conduct an omnibus test,
\begin{equation*}
H_0: \boldsymbol{\theta}_{(g)}=\boldsymbol{\theta}_0  \; {\mbox vs.} \;
H_1: \boldsymbol{\theta}_{(g)} \neq \boldsymbol{\theta}_0.
\end{equation*}
Here, $H_0$ indicates the situation when parameter $\boldsymbol{\theta}$ remains constant (that is, {\it homogeneity}) and $H_1$ corresponds to the situation of parameter instability (that is, {\it heterogeneity}) . The parameter instability tests utilize the following properties of score function under $H_0$:
\begin{itemize}\setlength{\itemsep}{-0.2em}
\item{A1:} $E_{H_0}[\mathbf{u}(\mathbf{y}_i, \boldsymbol{\theta}_0)]=0$;
\item{A2:} $\mbox{Var}_{H_0}[\mathbf{u}(\mathbf{y}_i, \boldsymbol{\theta}_0)]=\mathbf{J}(\boldsymbol{\theta}_0)=\mathbf{J}$;
\item{A3:} $\mathbf{u}(\mathbf{y}_i, \hat{\boldsymbol{\theta}}) |_{H_0}\rightarrow^d N[0, \hat{\mathbf{J}}] $,
\end{itemize}
where $\hat{\boldsymbol{\theta}}$ is the maximum likelihood estimate of $\boldsymbol{\theta}$ and $\hat{\mathbf{J}}=\mathbf{J}(\hat{\boldsymbol{\theta}})$. We discuss the instability test separately depending on whether the  partitioning  variable $X^G$ is  categorical or  continuous.
%
%
\subsection{Instability test with a categorical partitioning variable}\label{InstabCate}
It is straightforward to obtain a test for parameter instability using the properties A1--A3 when the partitioning variable, $X^G$, is categorical with a small number of categories (that is, $G \ll N$).  Since the score functions $\mathbf{u}(\mathbf{y}_i, \hat{\boldsymbol{\theta}})$ are independent, we have under $H_0$, the following quantity
\begin{equation*}
\chi^2_{cat}=\sum_{g=1}^G{\left[\sum_{i=1}^N{I(X_i^G=c_{(g)})\mathbf{u}(\mathbf{y}_i, \hat{\boldsymbol{\theta}})}\right]^{\top}\left[m_g\hat{\mathbf{J}}\right]^{-1}\left[\sum_{i=1}^N{I(X_i^G=c_{(g)})\mathbf{u}(\mathbf{y}_i, \hat{\boldsymbol{\theta}})}\right]}
\end{equation*}
is asymptotically distributed as $\chi^2$ with $(G-1)p$ degrees of freedom where $p$ is the dimension of $\boldsymbol{\theta}$. Here, $I(\cdot)$ is the indicator function. The reduction in $p$ degrees of freedom is due to the estimation of $p$ dimensional $\boldsymbol{\theta}$ from the data.
%
%
\subsection{Instability test with continuous partitioning variable}\label{InstabContinuous}

Our proposed instability test for continuous partitioning variable is based on score process. We begin by defining the following {\em score process}
\begin{equation*}
 \mathbf{W}_N(t, \boldsymbol{\theta}_0)=N^{-1/2}\sum_{i=1}^{M_g}{\mathbf{u}(\mathbf{y}_i, \boldsymbol{\theta}_0)} \qquad t\in [t_g, t_{g+1})
\end{equation*}
where $t_g=\dfrac{M_g}{N}$. Under $H_0$, using multivariate version of Donsker's theorem and Cram\'er-Wold theorem \citep[see e.g.][]{billingsley2009convergence}, it can be shown that
\begin{equation*}
 \mathbf{W}_N(t, \boldsymbol{\theta}_0) \rightarrow_d \mathbf{Z}(t)
\end{equation*}
where $\mathbf{Z}(t)$ is the zero-mean Gaussian process with $\mbox{cov}[\mathbf{Z}(t), \mathbf{Z}(s)]=\min(t,s)\mathbf{J}(\boldsymbol{\theta}_0)$. Since $\boldsymbol{\theta}_0$ is unknown in practice, we replace $\boldsymbol{\theta}_0$ by $\hat{\boldsymbol{\theta}}$ in score process 
\begin{equation*}
 \mathbf{W}_N(t, \hat{\boldsymbol{\theta}})=N^{-1/2}\sum_{i=1}^{M_g}{\mathbf{u}(\mathbf{y}_i, \hat{\boldsymbol{\theta}})}
\end{equation*}
\citet{hjort2002tests} has shown that the above estimated score process converges to Brownian Bridge process. We present this result as following theorem and the proof of the theorem is outlined in Appendix.
\begin{thm} \label{ThmCanM}
Let's define the standardized estimated score process as 
\begin{equation*}
 \mathbf{M}_N(t, \hat{\boldsymbol{\theta}}) =\hat{\mathbf{J}}^{-1/2} \mathbf{W}_N(t, \hat{\boldsymbol{\theta}}) 
\end{equation*}
Then under $H_0$,
\begin{equation*}
 \mathbf{M}_N(t, \hat{\boldsymbol{\theta}}) \rightarrow_d \mathbf{W}^0(t)
\end{equation*}
where $\mathbf{W}^0(t) = (W^0_1(t), \ldots, W^0_p(t))$ is a vector with $p$ independent standard Brownian Bridges as component processes. 
\end{thm}
Since the limiting distribution is the vector of independent Brownian Bridge process, individual  components of $\mathbf{M}_N(t, \hat{\boldsymbol{\theta}}) $ is distributed as a standard Brownian Bridge, $W^0(t)$. That is,
\begin{equation*}
M_N(t, \hat{\theta}_k) \rightarrow_d W^0(t)   \;\;\; k^{th}\;\; (k=1, \cdots, p)
\end{equation*}
The above weak convergence continues to hold for any `reasonable' functionals (including supremum) of  $M_N(t, \hat{\theta}_k)$  \citep[see e.g. ][pp 509, Theorem 1]{csorgo2002glimpse}. Therefore,
\begin{equation}
D_k\equiv \max_{0 \le t \le 1}{|M_N(t, \hat{\theta}_k)|} = \max_{1 \le j \le N-1}{ |M_N(t, \hat{\theta}_k)|} \rightarrow_d \max_{0 \le t \le 1}{|W^0(t)|}\equiv D \label{eq:D_k}
\end{equation}
 $D$ has known distribution function \citep{billingsley2009convergence}
\begin{equation*}
F_D(x)=1+2\sum_{l=1}^{\infty}{(-1)^l\exp{(-2\;l^2x^2)}}.
\end{equation*}
Although this expression involves an infinite series, this series converges very rapidly. Usually a few terms suffice for very high accuracy. This result can be used to formulate a test for instability of parameters at $\alpha$ level of significance as follows: (1) Calculate the value of the process $D_k$ for each parameter $k=1,\ldots,p$ and obtain the raw p-values. (2) Adjust the p-values according to a chosen multiple testing procedure. (3) Reject $H_0$ if the adjusted p-value for any of the processes, $D_k$, is less than $\alpha$.


%
\subsection{Instability test for multiple partitioning variables}\label{MultPartitioning}
In practice, we expect to have more than one partitioning variable. Let there be $S$ partitioning variables: $\{X^{G_1}_1, \ldots, X^{G_S}_S\}$. In that case we need to perform the instability test for each of the partitioning variables $X^{G_1}_1, \ldots, X^{G_S}_S$ subject to adjustment for multiplicity of type I errors. Let the p-values after multiplicity adjustment be $p_1, \ldots, p_S$, respectively and $p_{min}=\min{\{p_1, \ldots, p_S\}}$. Candidate partitioning variable with the smallest p-value ($p_{min}$) is chosen as a partitioning variable if $p_{min}$ is smaller than the nominal significance level. For further discussion please see Section \ref{TreeConstruction}.
%
%
\subsection{Power under the alternative hypothesis}
We consider the following form of Pitman's local alternatives in the vicinity of $H_0$
\begin{equation} \label{thetag}
\boldsymbol{\theta}_{(g)}=\boldsymbol{\theta}_0+\boldsymbol{\delta} \circ \mathbf{h}\Big(\frac{c_{(g)}}{c_{(G)}}\Big)\frac{1}{\sqrt{N}} + O\left(\frac{1}{N}\right)
\end{equation}
where $\boldsymbol{\delta}=(\delta_1, \ldots, \delta_p)^{\top}$ is the vector containing degrees of departure from the null hypothesis and $\mathbf{h}=(h_1, \ldots, h_p)^{\top}$ is the vector containing magnitudes of departure. The operation $\circ$ denotes the point-wise multiplication, i.e.,
\[
\boldsymbol{\delta} \circ \mathbf{h}\Big(\frac{c_{(g)}}{c_{(G)}}\Big)=\left[\delta_1 h_1\Big(\frac{c_{(g)}}{c_{(G)}}\Big), \ldots, \delta_p h_p\Big(\frac{c_{(g)}}{c_{(G)}}\Big)\right]^{\top}
\]

\begin{thm} \label{ThmChi}
Under \eqref{thetag}, the limiting distribution for the $\chi^2_{cat}$ is a non-central chi-square distribution
\[
\chi^2_{cat}   \longrightarrow_d \chi'^2\left[(G-1)p,\;\; \sum_{g=1}^G{\lambda^2_g}\right]
\]
where $\lambda_g= \mathbf{J}\cdot m_g\mathbf{h}\Big(\dfrac{c_{(g)}}{c_{(G)}}\Big)\cdot\dfrac{1} {\sqrt{N}}$
\end{thm}

\begin{thm} \label{ThmMn}
Under \eqref{thetag}, the limiting distribution for the canonical monitoring process is as follows
\[
\mathbf{M}_N(t, \hat{\boldsymbol{\theta}})   \longrightarrow_d \mathbf{J}^{1/2} \cdot t_g \cdot \boldsymbol{\delta} \circ (\bar{\mathbf{h}}_g - \bar{\mathbf{h}}) +\mathbf{W}^0(t)
\qquad t\in [t_g, t_{g+1})
\]
where, $\bar{\mathbf{h}}_g= \dfrac{1}{M_g}\sum_{j=1}^{g}{m_j\mathbf{h}\Big(\dfrac{c_{(j)}}{c_{(G)}}\Big)}$ and $\bar{\mathbf{h}}=\bar{\mathbf{h}}_G$
\end{thm}

Proofs of these theorems are provided in the Appendix.

\section{Longitudinal Regression Tree}\label{TreeConstruction}
\subsection{LongCART Algorithm}\label{Algorithm}
Smaller p-values from the instability test indicate greater  evidence towards instability. Intuitively, splits in the tree should be based on the partitioning variable that shows higher evidence towards instability of the parameters.  Therefore, we propose the following algorithm in order to construct a regression tree for longitudinal data.
\begin{description}
\item {{\bf Step 1.}} Perform the instability test for each partitioning variable separately at a prespecified level of significance $\alpha$. The level of significance for performing instability test is subject to adjustment for multiple comparisons in order to maintain the level of type I error.
\item {{\bf Step 2.}} Stop if no partitioning variable is significant at level $\alpha$. Otherwise, choose the partitioning variable with the smallest p-value and proceed to step 3.
\item {{\bf Step 3.}} Consider all cut-off points of the chosen partitioning variable. At each cut-off point, calculate the improvement in the goodness of fit criterion (e.g., deviance). With $X^G$ as the chosen partitioning variable, the improvement in goodness of fit criterion can be obtained at the cut-off point $c_{(g)}$ in the following steps:
	\begin{description}
	\item {{\bf  a.}} Split the data in two parts. One group will include the observations from subjects with $X^G\leq c_{(g)}$ and the other group will have the observations from subjects with $X^G> c_{(g)}$.
	\item {{\bf b.}} Fit the longitudinal model on (i) all the individuals in the node, (ii) the individuals with $X^G\leq c_{(g)}$ and (iii) the individuals with $X^G> c_{(g)}$. Calculate the goodness of fit criterion from each of these three models. Call them as $\mbox{GOF}_{all}$, $\mbox{GOF}_{I}$ and $\mbox{GOF}_{II}$, respectively.
	\item {{\bf c.}} Calculate the improvement in goodness of fit criterion as $\mbox{GOF}_{I} + \mbox{GOF}_{II} - \mbox{GOF}_{all}$.
	\end{description}
\item {{\bf Step 4.}} Choose the cut-off value that provides the maximum improvement in goodness of fit criterion and use this cut-off for binary splitting.
\item {{\bf Step 5.}} Follow the Steps 1-4 for each non-terminal node.
\end{description}
The above strategy for construction of regression tree with longitudinal data has two major advantages over the currently existing algorithms. First, the decision about further splitting at each node is taken controlling type I error. Second, there are huge savings in computation time as we are evaluating the improvement in selected goodness of fit criterion at the cut-off points of the chosen partitioning variable only.
%
\subsection{Improvement}
A measure of improvement due to regression tree can be provided in terms of likelihood function based criterion. For example, Akaike Information criterion (AIC) for a tree $T$ can be obtained as
\[
\mbox{AIC}_T=2\sum_{k=1}^{|T|}{l_k} - 2 \cdot |T| \cdot p
\]
where $|T|$ denotes the number of terminal nodes in $T$, $l_k$ is the log-likelihood in $k$th terminal node and $p$ is the number of estimated parameters in each node. If we denote the AIC obtained from the traditional linear mixed effects model at root node (that is, common parametric form for covariates and time for the entire population)  by $\mbox{AIC}_0$, the improvement due to regression tree can be measured as
\[
\mbox{Improvement }(T) = \mbox{AIC}_T-\mbox{AIC}_0
\]
Since the overall model fitted to all the data is nested within the regression tree based model, a likelihood ratio test or test for deviance can be constructed as well to evaluate the overall significance of a given regression tree.
\subsection{Pruning}
The improvement in regression tree comes at a cost of adding complexity to the model. If we can summarize complexity of a tree by number of terminal nodes, the cost adjusted AIC of a regression tree $T$ can be defined as follows
\[
\mbox{AIC}_T(\gamma)=\mbox{AIC}_T-\gamma (|T|-1), \;\;\;\; \gamma>0
\]
where $\gamma$ be the  {\it average complexity} for each terminal node. Hence, the tree $T$ will be selected if
\[
\mbox{AIC}_T-\gamma (|T|-1)>\mbox{AIC}_0
\]
\begin{equation}
\mbox{or} \qquad \gamma<\frac{\mbox{AIC}_T-\mbox{AIC}_0}{|T|-1} \equiv \gamma_T  \label{eq:AvgCost}
\end{equation}
That is, the tree $T$ will be chosen as long as $\gamma_T$ does not exceed  some pre-set level of {\it average complexity}, $\gamma_0$; otherwise, we have to  prune the tree $T$ to bring $\gamma_T$ below $\gamma_0$.
%
%
\section{Simulation}\label{SimulationRegTree}
We have explored the performance of instability test for continuous partitioning variables and the performance of proposed LongCART algorithm as a whole through simulation studies. 
%
\subsection{Performance of instability test with continuous partitioning variable} \label {SimulationRegTree1}
Let $X^G$ be continuous partitioning variable with ordered cut-off points as $c_{(1)}\leq \ldots \leq c_{(G)}$. We first investigated the size of the test and then obtained the size-corrected power.
%
\subsubsection{Size of the test}\label{sizeoftest}
\begin{table}
\begin{center}
\caption{Size of proposed parameter instability test for continuous partitioning variable via simulation as discussed in Section~\ref{sizeoftest}. The results were summarized based on $10,000$ simulations for various nominal levels of type I error ($\alpha$) and sample size ($N$). The critical values ($D_{\alpha}$) from the true limiting distribution of test statistic $D_k$  (see Eq.~\ref{eq:D_k}) is also provided for each $\alpha$. For each $N$ and $\alpha$, the simulation results have been summarized by (a) percentage of rejection (to be compared with $\alpha$) and (b) observed $(1-\alpha)100^{th}$ percentile of $D_k$ (to be compared with $D_{\alpha}$). The propose parameter instability test  seems to be conservative; however, the size of the test approaches to nominal level with the increase in $N$.}
\label{Sim1Size}
\vspace{0.1in}
(a) Percentage of rejection

\begin{tabular}
{c|r|r|r|r|r}
\hline
  & \multicolumn{5}{c}{$N$}\\
\hline
$\alpha (\%)$ & $50$ & $100$   & $200$ & $500$ & $1000$ \\
\hline
 \hline
1.25 &0.54 & 0.56& 0.89 &1.02 & 0.95\\
1.67 & 0.75 & 0.85& 1.10 &1.33& 1.29\\
2.50 & 1.20 & 1.46& 1.77 &2.04 & 1.94\\
5.00 & 2.78 & 3.35 & 3.48 & 4.07&4.19 \\
10.00&5.66 &7.14 & 7.19&8.37 &8.53\\
20.00 & 13.05& 14.73 & 15.83&16.97 & 17.14\\
 \hline
\end{tabular}

\vspace{0.4in}
(b) Observed $(1-\alpha)100^{th}$ percentile of $D_k$\\

\begin{tabular}
{c|r|r|r|r|r|r}
\hline
&& \multicolumn{5}{c}{$N$}\\
\hline
$\alpha (\%)$ & $D_{\alpha}$ & $50$ & $100$   & $200$ & $500$ & $1000$ \\
\hline
 \hline
1.25& {\bf 1.5930}&  1.4760& 1.4938 & 1.5366& 1.5643& 1.5504\\
1.67& {\bf 1.5472} & 1.4447 & 1.4532& 1.4891& 1.5147 & 1.4986\\
2.50&{\bf 1.4802}& 1.3722 & 1.3998& 1.4180 & 1.4392& 1.4412\\
5.00& {\bf 1.3581}& 1.2497 & 1.2924 & 1.2934 & 1.3154 &1.3287\\
10.00&{\bf 1.2238} & 1.1236& 1.1585& 1.1629 & 1.1901&1.1857\\
20.00&{\bf 1.0728}& 0.9859& 1.0045& 1.0194& 1.0350&1.0373\\
 \hline
\end{tabular}

\end{center}
\end{table}

In order to examine the size of the test we have considered  a longitudinal model with single mean parameter. We generated observations for $N$ subjects at $t=0, 1, 2, 3$  from the following model
\begin{equation}\label{StableEq}
X^G=c_{(g)}: \;\;y_{it} =\beta_{0} + b_i + \epsilon_{it}
\end{equation}
with $\beta_0=2$, $b_i \sim N(0, 0.5^2) $ and  $\epsilon_{it} \sim N(0, 0.2^2)$. The observations for $X^G$ were generated for each simulation separately from uniform(0,300). For each $N$,  $10,000$ Monte-Carlo samples were generated and the test statistic $D_k$ (see Eq. \eqref{eq:D_k}) was calculated for each sample separately. The null hypothesis of parameter stability is rejected at $\alpha$\% level of significance when $D_k$ exceeds the $(1-\alpha)\times 100$th percentile of the limiting distribution.

The observed percentiles and the percentage of rejected null hypotheses are summarized in Table~\ref{Sim1Size}. We can make following observations: 1) the type I error of test does not exceed the nominal level,  2) the size of the test approaches to the desired significance level $\alpha$ with the increase in the sample size $N$, and 3) the test is under-sized for smaller sample sizes. The severe problem with the size of the test for smaller sample size can be explained as follows. Calculation of test statistic, $D_k$, involves $\sigma^2$ and $\mathbf{V}_i$. However, in practice, the true values of  $\sigma^2$ and $\mathbf{V}_i$ are unknown and we replace them by their estimates. A consistent estimator (e.g. ML- or REML-based) approaches the true value with an increasing sample size. However, the estimates might be biased for smaller sample sizes. To be precise, for smaller sample size,  $\sigma^2$ and $\mathbf{V}_i$ may remain underestimated and this leads to smaller value of $D_k$ which in turn results in a smaller size of the test. However, bias in estimation of $\sigma^2$ and $\mathbf{V}_i$ fades away with the increase in $N$ and this increases the size of the test. We observe this trend in Table~\ref{Sim1Size} as the size of test approaches the nominal level of type I error with the increase in sample size. However, the size of test remains smaller than nominal level even for the reasonably large $N$. The reduced size has been also reported in other tests based on the Brownian Bridge process. For example, Kolmogorov Smirnov test for normality (which also uses the Brownian Bridge as limiting distribution) is conservative \citep{lilliefors1967kolmogorov, massey1951kolmogorov, birnbaum1952numerical}.  As $N$ exceeds 500, the size of the test is  close to the nominal level of significance. As a remedy for smaller sample sizes, one might consider using a liberal $\alpha$ level or small sample distribution for $D_k$ obtained through simulation.

\subsubsection{Power}\label{InstabilityPower}
We generated observations for $N$ subjects at $t=0, 1, 2, 3$  from the following model to evaluate performance of instability test for $X^G$
\[
X^G=c_{(g)}: \;\;y_{it} =\beta_{0(g)} + \beta_{1(g)} t  + b_i + \epsilon_{it},
\]
\[
\beta_{0(g)} = \beta_0 \qquad
\beta_{1(g)} = \beta_1+\delta \cdot\frac{c_{(g)}}{c_{(G)}}
\]
We set $\beta_0=1$ and $\beta_1=2$. $b_i$,  $\epsilon_{it}$ and $X^G$ were generated similarly as before in Section \ref{sizeoftest}. In this simulation, the parameter $\beta_1$ is not stable unless $\delta=0$. We dealt with two parameters: $\beta_0$ and $\beta_1$, thus we will have two Brownian bridge processes. We adjusted the p-values according to the Hochberg's step-up procedure \citep{hochberg1988sharper}. We  chose Hochberg's step-up procedure because it is relatively less conservative than the Bonferroni procedure \citep{hochberg1987multiple}. However, in principle, any multiple comparison procedure can be applied here.
\begin{table}
\begin{center}
\caption{Power (\%) of parameter instability test with continuous partitioning variable obtained  in the simulation described in Section \ref{SimulationRegTree1}. Numbers corresponding to $\beta_1$ and $\beta_0$ represent the percentages of rejection associated with parameter instability for $\beta_1$ and $\beta_0$, respectively. The `Overall' figures represent the percentage of at least one rejection out of the two.} \label{Sim1TreeSummary}
\begin{tabular}
{r|c|r|r|r|r|r|r}
\hline
 &  & \multicolumn{6}{c}{\% of rejection}\\
\cline{3-8}
 &  & \multicolumn{6}{c}{$\delta$}\\
\cline{3-8}
    & Parameter & &&&&&\\
 N &  instability test & $0$   & $0.25(-0.25)$ & $0.50(-0.50)$ & $0.75(-0.75)$ & $1.00(-1.00)$  &  $1.2(-1.2)$\\
\hline
      & $\beta_1$ &1.4&1.4(1.4)&1.6(1.6)&1.9(1.9)&2.3(2.3)&2.4(2.3)\\
50   & $\beta_0$ &1.6&4.4(4.3)&16.9(16.6)&41.9(42.0)&70.2(70.6)&86.9(87.0)\\
      &  Overall     &2.9&5.6(5.5)&17.9(17.6)&42.6(42.5)&70.5(70.8)&87.0(87.1)\\
\hline
      & $\beta_1$ &1.5&1.6(1.6)&2.0(2.1)&2.5(2.6)&3.0(3.0)&3.2(3.2)\\
100  & $\beta_0$ &1.7&5.2(5.3)&18.7(19.7)&44.4(46.0)&72.9(73.9)&88.9(89.0)\\
      &  Overall     &3.1&6.6(6.7)&19.8(20.8)&45.0(46.6)&73.1(74.2)&89.0(89.1)\\
\hline
      & $\beta_1$ &1.8&1.9(1.8)&2.2(2.2)&2.7(2.7)&3.3(3.3)&3.5(3.4)\\
200  & $\beta_0$ &1.9&5.6(5.3)&20.7(19.8)&47.5(46.8)&75.7(75.2)&90.1(89.8)\\
      &  Overall     &3.6&7.4(6.8)&21.9(21.0)&48.2(47.4)&76.0(75.4)&90.6(89.9)\\
\hline
      & $\beta_1$ &2.1&2.1(2.2)&2.7(2.5)&3.2(3.2)&3.6(3.7)&3.9(4.0)\\
500  & $\beta_0$ &1.8&6.1(6.0)&21.4(20.1)&48.1(48.2)&76.6(76.6)&91.1(91.1)\\
      &  Overall    &3.7&7.8(7.8)&22.8(22.2)&48.8(49.1)&77.0(77.0)&91.3(91.2)\\
 \hline
\end{tabular}
\end{center}
\end{table}
The results based on 10,000 simulation are displayed in Table \ref{Sim1TreeSummary}. As the absolute value of $\delta$ deviates from zero, the power increases. The power of test is close to 80\% and approaching the 90\% mark as $|\delta|> 1$. The sign of $\delta$ does not influence the power of the test. Sizes of the test are very much in agreement with the first simulation study. As observed previously, the test is mildly conservative in the current simulation scenario as the observed level of type I error is consistently slightly below the nominal value $\alpha=0.05$.

\subsection{Performance of regression tree for longitudinal data}\label{SimulationRegTree2}
\begin{figure}
\centerline{\includegraphics [angle=0,width=120mm, height=60mm]{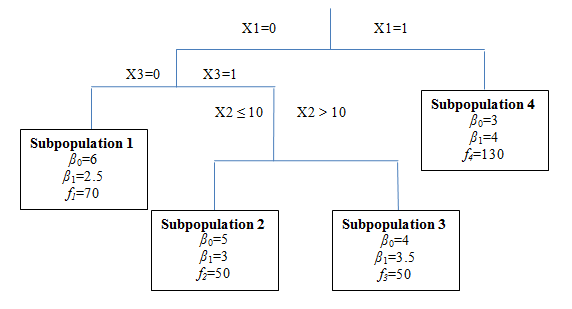}}
\caption{True tree structure for the simulation described in section~\ref{SimulationRegTree2}. In $r^{th}$ subgroup, $f_r$ observations were generated according to Eq.~\eqref{eq:longtree_sim2} with specified $\beta_0$ and $\beta_1$. } \label{Sim2TrueTree}
\end{figure}

In this simulation, our goal is to assess the improvement in estimation due to LongCART algorithm when the population under consideration is truly heterogeneous. We have simulated observations for  $N=300$ subjects and these subjects come from one of the four different subgroups. Description of these subgroups is displayed in the form of a tree structure in Figure \ref{Sim2TrueTree}.  The subgroups can be defined in terms of the partitioning variables $X_1$, $X_2$ and $X_3$. In $r$th subgroup ($r=1,\ldots, 4$), the values for continuous response variable $y$ were generated at $t=0, 1, 2, 3$ according to following model:
\begin{equation}
y_{it} =\beta_{0r} + \beta_{1r} t  + b_i + \epsilon_{it}; \;\;\;\;\; i=1,\ldots, f_r
\label{eq:longtree_sim2}
\end{equation}
where  $b_i \sim N(0, 4)$  and $\epsilon_{it} \sim N(0, 1)$. As displayed in Figure \ref{Sim2TrueTree}, the true values of $\beta_1$ were set at $2.5$, $3.0$, $3.5$ and $4.0$  and for $\beta_0$, the true values were set at $6$, $5$, $4$ and $3$, for the four subgroups, respectively. Further, observations were generated for $f_1=70$ individuals in subgroup 1, $f_2=50$ individuals in subgroup 2, $f_3=50$ individuals in subgroup 3, and $f_4=130$ individuals in subgroup 4. In order to study the performance of our algorithm constructing the longitudinal regression tree, we calculated the mean absolute deviation (MAD) in $\beta_0$ and $\beta_1$ in $r$th subgroup for each simulation as defined below
\[
\mbox{MAD}(\hat{\beta}_{0r})=\frac{1}{ f_r}\sum_{j \in S_r}{|\beta_{0r}-\hat{\beta}_{0j}|}
\qquad
\mbox{MAD}(\hat{\beta}_{1r})=\frac{1}{ f_r}\sum_{j \in S_r}{|\beta_{1r}-\hat{\beta}_{1j}|},
\]
where $\beta_{0r}$ and $\beta_{1r}$ are the true values of $\beta_{0}$ and $\beta_{1}$ in the $r$th subgroup and  $\hat{\beta}_{0j}$ and $\hat{\beta}_{1j}$ are the corresponding estimates for the $j$th individual applyig longitudinal tree and then fitting mixed model in each subgroup. $S_r$ is the set of indices for all individuals  in the $r$th subgroup while $f_r$ denotes their number.

\begin{table}
\begin{center}
\caption{Description of the mixed models used in section \ref{SimulationRegTree2} for the comparison with LongCART algorithm (Model 1). All models include random intercepts to account for the subject-specific effects.} \label{Sim2Models}
\begin{tabular}
{l|l}
\hline
 & Predictors\\
\hline
Model 2 & $t$\\
Model 3 & $t$, $X_1$, $X_2$, $X_3$\\
Model 4 & $t$, $X_1$, $X_2$, $X_3$, $X_1X_2$, $X_1X_3$, $X_2X_3$ \\
Model 5 & $t$, $X_1$, $X_2$, $X_3$, $X_1X_2$, $X_1X_3$, $X_2X_3$, $X_1X_2X_3$ \\
Model 6 & $t$, $X_1$, $X_2$, $X_3$, $tX_1$, $tX_2$, $tX_3$\\
Model 7 & $t$, $X_1$, $X_2$, $X_3$,  $X_1X_2$, $X_1X_3$, $X_2X_3$, $tX_1$, $tX_2$,  $tX_3$, \\
        & $tX_1X_2$, $tX_1X_3$, $tX_2X_3$\\
Model 8 & $t$, $X_1$, $X_2$, $X_3$,  $X_1X_2$, $X_1X_3$, $X_2X_3$, $X_1X_2X_3$, $tX_1$, $tX_2$,  $tX_3$,  \\
        & $tX_1X_2$, $tX_1X_3$, $tX_2X_3$, $tX_1X_2X_3$\\
\hline
\end{tabular}
\end{center}
\end{table}
\begin{table}
\begin{center}

\caption{Summary of the results for the simulation described in section \ref{SimulationRegTree2}} \label{Sim2TreeSummary}
\begin{tabular}
{l|r|r|r|r|r|r|r|r|r|r|r}
\hline
 &  &\multicolumn{2}{c|}{Subpop 1} &  \multicolumn{2}{c|}{Subpop 2} &  \multicolumn{2}{c|}{Subpop 3} &  \multicolumn{2}{c|}{Subpop 4}&  \multicolumn{2}{c}{Overall}\\
\hline
&Par. &mad0&mad1&mad0&mad1&mad0&mad1&mad0&mad1&mad0&mad1\\
\hline
Model 1 &  8$^{\star}$ & 0.291 & 0.083 & 0.309 & 0.093 & 0.315 & 0.088 & 0.159 &  0.033 &0.241 & 0.064\\
\hline
Model 2 &  2 & 1.802 & 0.899 &  0.802 & 0.399  &  0.204 & 0.101  &  1.198 & 0.601 &1.107 & 0.554\\
Model 3 &  5 & 1.439  &0.899  & 0.615 &  0.399  & 0.320  & 0.101 &  0.893  &  0.601 &0.879 & 0.554\\
Model 4 &  8 & 1.345 & 0.899 & 0.641  & 0.399  &  0.277 & 0.101 &  0.895 & 0.601 &0.855 &0.554\\
Model 5 &  9 & 1.345  & 0.899   & 0.628  & 0.399   & 0.281 & 0.101 &  0.896 & 0.601 & 0.854 &0.554 \\
\hline
Model 6 &  8 & 0.404 &  0.185  &  0.202  & 0.061 &   0.585  & 0.292 &  0.432 & 0.200 &0.413 & 0.189\\
Model 7 & 14 & 0.286  & 0.070  & 0.306 & 0.114  & 0.300 & 0.105 &  0.291 & 0.074 &0.294 &0.085\\
Model 8 & 16 & 0.294& 1.822 &   0.309&  1.225 &  0.299 &  2.181 & 7.089 & 2.256 & 3.242 &1.970\\
 \hline

\multicolumn{12}{l}{mad0= Average $\mbox{MAD}(\hat{\beta}_0)$, mad1= Average $\mbox{MAD}(\hat{\beta}_1)$, Par.: No.of parameters.} \\
\multicolumn{12}{l}{Model 1: Subgroups are extracted using LongCART algorithm and mixed model with  time}\\
\multicolumn{12}{l}{slope and random intercept fitted separately in each subgroup. }\\
\multicolumn{12}{l}{Model 2 - 8: Description is given in Table \ref{Sim2Models}}\\
\multicolumn{12}{l}{$^{\star}$ - In Model 1, $81\%$ of time regression tree with 4 subgroups were extracted.}
\end{tabular}
\end{center}
\end{table}

The simulation results are summarized in Table \ref{Sim2TreeSummary} based on 1000 simulations in each case. In each simulation, regression tree was constructed with the following specifications: (1) the overall significance level of instability test was set at 5\%, (2) minimum node size for further split was set at 40, and (3) minimum terminal node size was set at 20. Recall that we are considering four subgroups in the current simulation. The LongCART algorithm extracted exactly four subgroups in 81\% of the cases. \textcolor{black}{Five subgroups were extracted in 16\% of the cases and in these trees we observed a split in subgroup 4 which was not present in the true tree (see Figure~\ref{Sim2TrueTree}).} There were only 1.3\% instances when 3 subgroups and 1.6\% instances when 6 subgroups were extracted.

For the comparison purposes, we considered seven linear mixed models (Model 2 -- Model 8). These models are described in Table \ref{Sim2Models}. The application of the LongCART algorithm (Model 1) shows comparatively larger improvements in the estimation of the coefficients in all four subgroups. Both the $\mbox{MAD}(\hat{\beta}_0)$ and $\mbox{MAD}(\hat{\beta}_1)$ were considerably smaller in Model 1 compared to the Models 2---8. The improvement in estimation of coefficients in regression tree was attributed to its ability to extract homogeneous subgroups and then fitting mixed model separately within each group. On the contrary, Models 2---8 assume either additive (Models 2---3) or an interaction (Models 4---8) mixed effects model for the entire population assuming parametric form for both covariates and time. These models do not capture the complexity for the heterogeneous subgroups and overestimate it for the homogeneous subgroups.

Inclusion of the interaction terms in the model does not necessarily take into account subgroup heterogeneity in the presence of continuous partitioning variable. For example, in Models 4 and 5 common slope is assumed for the entire population, but they include interaction terms for the baseline effects; still, the absolute deviation in estimating $\beta_0$ is almost $2.5$ times higher compared to that of a longitudinal tree. Similarly, Models 6 -- 8 include interaction terms for both baseline and longitudinal effects, but again the absolute deviations in estimating $\beta_0$ and $\beta_1$ are higher compared to what we have obtained with the longitudinal regression tree.

Model 6 including the interaction terms with $t$ and the partitioning variables is probably the most commonly used model in practice. However, the application of the LongCART algorithm  offers a considerable improvement in the estimation compared to Model 6. Models 6 -- 7 provide some improvement over regression tree in some of the subgroups. However, these improvements are comparatively rare and largely influenced by the fact how the subgroups are defined. We would close this section pointing out, apart from providing improvement in estimation, the LongCART algorithm also identifies the meaningful subgroups defined by the partitioning variables which would  remain unidentified otherwise.
%
\section{Application}\label{TreeApplication}
	\textcolor{black}{We applied the  LongCART algorithm to study the changes in concentration of a metabolite {\it choline} in gray matter region of the brain among HIV patients enrolled in the HIV Neuroimaging Consortium (HIVNC) study \citep{gongvatana2013progressive}. At the time of enrollment, patients were on stable FDA approved antiretroviral  therapy. Concentrations of choline were obtained via magnetic resonance spectroscopy (MRS). Choline is considered to be a marker of brain inflammation. It has been found in previous studies that the concentrations of choline were elevated in all three brain regions among HIV patients  \citep{chang2002relationships}. We considered a total of  $\sum_{i=1}^N{n_i}=780$ observations from $N=239$ subjects. All the observations were within 3 years from baseline. The number of observations per subject ranged from 2 to 6 with median equal to 3. We estimated the overall significant decrease of 0.077 AU per year (p-value=0.003) in choline concentration suggesting overall beneficial effect of antiretroviral therapy.}


For the construction of regression tree we used baseline measurements of several clinical and demographic variables including sex, race, education, age, current CD4 count, nadir CD4 count, duration of HIV infection, duration of antiretroviral (ARV) treatment, duration of highly active antiretroviral therapy (HAART), plasma HIV RNA count, antiretroviral CNS penetration-effectiveness (CPE) score and AIDS dementia complex (ADC) stage as partitioning variables. In each node we consider fitting the following model separately
\begin{equation}
y_{it}=\beta_0 + \beta_1 t +b_i +\epsilon_{it}
\label{LongCARTAppEq}
\end{equation}
\begin{figure}
\centering
\caption{{\it Left panel.} Longitudinal regression tree obtained via LongCART algorithm for longitudinal change in {\it choline} concentration as discussed in Section~\ref{TreeApplication}. The p-value in each node corresponds to the estimate of the slope $\beta_1$. {\it Right panel.} Estimated linear trajectory for longitudinal change within each subgroup obtained via fitting mixed effect model of the form Eq.~\eqref{LongCARTAppEq}. This regression tree suggests duration of ARV treatment and HAART are significant determinants for longitudinal change of choline.} \label{ChoMFCTree}
\begin{subfigure}{.7\textwidth}
  \centering
  \includegraphics[angle=0,width=100mm, height=60mm]{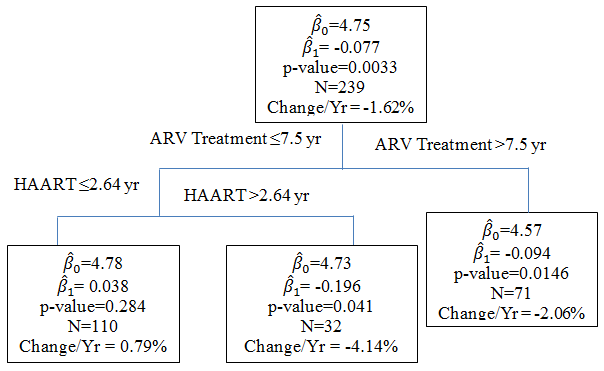}
\end{subfigure}%
\begin{subfigure}{.3\textwidth}
  \centering
  \includegraphics[angle=0,width=60mm, height=60mm]{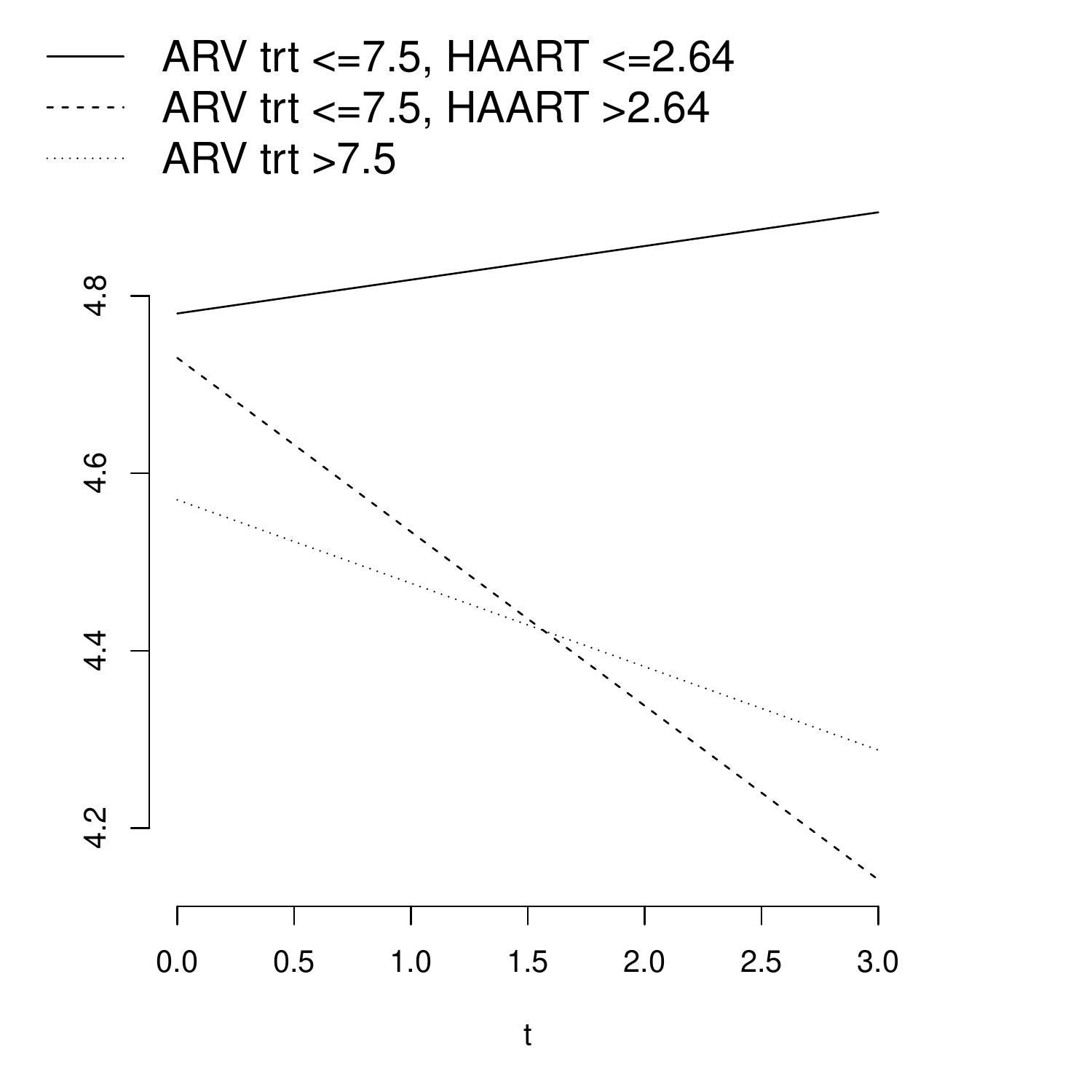}
\end{subfigure}
\end{figure}
\textcolor{black}{where $y_{it}$ indicates the measurement of the  concentration of choline from the $i$th individual at time $t$ (in years) and $b_i$ is the subject specific intercept. It was assumed that $b_i$ and $\epsilon_{it}$ are independently and normally distributed with mean equal to zero. LongCART algorithm  was applied with the following specifications: (1) the significance level for individual instability test was set to 5\%, (2) the minimum node size for further split was set to 50, and (3) the minimum terminal node size was set to 25. }

\textcolor{black}{Figure \ref{ChoMFCTree} displays the estimated longitudinal regression tree with the estimates of $\beta_0$ and $\beta_1$ for each terminal node or subgroup and the plot of estimated linear trajectories within each subgroup.} \textcolor{black}{Duration of ARV treatment (p-value=0.004) and HAART (p-value=0.004) seem to influence the change in concentration of choline over time. Improvement in deviance due to application of LongCART algorithm was 519 (log-likelihoods were $-1427$ vs. $-1687$; with 4 degrees of freedom).    ARV treatment for over 7.5 years not only helped to reduce baseline concentration of choline, but also resulted in a significant decrease of 0.094 per year (p-value=0.015). A higher baseline value of choline concentration was observed among those who received ARV treatment for at most 7.5 years; however, a longer period of HAART therapy in them led to significant decrease of 0.196 per year (p-value=0.041) in concentration over time. We did not observe any decrease among those who received ARV treatment for less than 7.5 years and HAART therapy for 2.64 years.}

\textcolor{black}{In summary, both the longer duration of ARV treatment and HAART resulted in reduction of choline concentration. However, the rate of reduction is almost double (4.14\% vs 2.06\%) when patients were on HAART compared to only ARV treatment (see Figure~\ref{ChoMFCTree}). This suggests that both ARV treatment and HAART are effective in controlling brain inflammation via reducing choline concentration. 
Finally, all these interpretable subgroups along with a significant improvement in overall model fit suggests underlying heterogeneity in the population in terms of longitudinal change in choline concentration. Thus considering a traditional linear mixed effects model for the entire population is not defensible.}


\section{Discussion}\label{sec:disc}
\textcolor{black}{The longitudinal profile in a population may be influenced by several baseline characteristics. This may be true both in observational studies and clinical trials.  The most common strategy to incorporate the effect of baseline variables in a traditional linear mixed effects model is to include these baseline characteristics and their interactions with the time-varying variables as covariates in the model.   However, this approach has its own limitations as discussed in Section \ref{sec:intro}. Longitudinal trees, i.e. regression trees for longitudinal data, are extremely useful to identify the heterogeneity in longitudinal  trajectories in a given population in a nonparametric way. We have proposed LongCART algorithm for the construction of longitudinal tree which firstly, controls type I error at the time of taking decision about splitting at each node. Secondly, LongCART algorithm reduces the computation time substantially as we first choose the partitioning variable and then evaluate the goodness of fit criterion at all cut-off points of the selected partitioning variable only.  Both the instability test and the LongCART algorithm discussed in this paper are based on the score process.  Therefore, we can extend the scope of LongCART algorithm to other scenarios as long as we can obtain (or approximate) an expression for the score function and the Hessian matrix in a tractable form including the generalized linear mixed effects model (GLMM) and multiple response variables setting.}

\appendix

\section{Proofs}\label{app}

\subsection{Proof of Theorem \ref{ThmCanM}}

\begin{proof}
Under $H_0$, by applying Taylor series expansion
\begin{equation*}
 \mathbf{W}_N(t, \hat{\boldsymbol{\theta}}) \doteq \mathbf{W}_N(t, \boldsymbol{\theta}_0) - t\;\mathbf{W}_N(1, \boldsymbol{\theta}_0)
\end{equation*}
where $A_n \doteq B_n$ means that $A_n - B_n$ tends to zero in probability. In the case of linear mixed effects models, this relationship is exact as the second derivative of the score function is equal to 0. That is, $\mathbf{W}_N(t, \hat{\boldsymbol{\theta}}) = \mathbf{W}_N(t, \boldsymbol{\theta}_0) - t\;\mathbf{W}_N(1, \boldsymbol{\theta}_0)$. Consequently,
\begin{equation*}
 \mathbf{W}_N(t, \hat{\boldsymbol{\theta}}) \rightarrow_d \mathbf{Z}(t) - t\cdot \mathbf{Z}(1) \equiv \mathbf{Z}^0(t)
\end{equation*}
The limit process $\mathbf{Z}^0(t)$ is a $p$-dimensional mean zero Brownian Bridge process with covariance function $\mbox{cov}[\mathbf{Z}^0(t), \mathbf{Z}^0(s)]=s(1-t)\mathbf{J}(\boldsymbol{\theta}_0)$ for $s<t$. Therefore, under $H_0$
\begin{equation*}
 \mathbf{M}_N(t, \hat{\boldsymbol{\theta}}) =\hat{\mathbf{J}}^{-1/2} \mathbf{W}_N(t, \hat{\boldsymbol{\theta}}) \rightarrow_d \mathbf{W}^0(t)
\end{equation*}
where $\mathbf{W}^0(t) = (W^0_1(t), \cdots, W^0_p(t))$ is a vector with $p$ independent standard Brownian Bridges as component processes. 
\end{proof}

\subsection{Proof of Theorem \ref{ThmChi}}

\begin{proof}
Using Taylor series expansion we can write
\[
f(\mathbf{y}, \boldsymbol{\theta}_{(g)}) \doteq f(\mathbf{y}, \boldsymbol{\theta}_0)\left\{1+\mathbf{u}(\mathbf{y},\boldsymbol{\theta}_0)^{\top}\boldsymbol{\delta} \circ \mathbf{h}\Big(\frac{c_{(g)}}{c_{(G)}}\Big)\frac{1}{\sqrt{N}}\right\}
\]
Consequently,

\begin{eqnarray}
E_{\boldsymbol{\theta}_g}[\mathbf{u}(\mathbf{y}, \boldsymbol{\theta}_0)] & = & \int{u(\mathbf{y},\boldsymbol{\theta}_0)f(\mathbf{y},\boldsymbol{\theta}_{(g)})dy}
=E_{\boldsymbol{\theta}_0}[\mathbf{u}(\mathbf{y}, \boldsymbol{\theta}_0)]+\mathbf{J} \cdot \boldsymbol{\delta} \circ \mathbf{h}\Big(\frac{c_{(g)}}{c_{(G)}}\Big)\frac{1}{\sqrt{N}} \nonumber \\
& =  & \mathbf{J} \cdot \boldsymbol{\delta} \circ \mathbf{h}\Big(\frac{c_{(g)}}{c_{(G)}}\Big)\frac{1}{\sqrt{N}} \label{ExpH1}
\end{eqnarray}
It can be shown that
\begin{equation}
\mbox{cov}_{H_1}[\mathbf{W}_N(t, \boldsymbol{\theta}_0)] = \mbox{cov}_{H_0}[\mathbf{W}_N(t, \boldsymbol{\theta}_0)]  + O\left(\frac{1}{N}\right)\doteq \mathbf{J} \label{VarH1}
\end{equation}
Proof of Theorem \ref{ThmChi} follows from the definition of non-central chi-square distribution.
\end{proof}

\subsection{Proof of Theorem \ref{ThmMn}}
\begin{proof}
Using \eqref{ExpH1} and \eqref{VarH1},
\[
E_{H_1}[\mathbf{W}_N(t, \boldsymbol{\theta}_0)]
       =\mathbf{J} \frac{1}{N}\sum_{i=1}^{M_g}{\delta \circ \mathbf{h}\Big(\frac{c_{(g)}}{c_{(G)}}\Big)}
       =\mathbf{J} \cdot t_g \cdot \boldsymbol{\delta} \circ \bar{\mathbf{h}}_g
\qquad t\in [t_g, t_{g+1})
\]

This time using the FCLT along with Cramer-Wold device we can show that
\[
 \mathbf{W}_N(t, \boldsymbol{\theta}_0) \longrightarrow_d \mathbf{J} \cdot t_g \cdot \delta \circ \bar{\mathbf{h}}_g + \mathbf{Z}(t) \qquad t\in [t_g, t_{g+1})
\]
Therefore, for $t\in [t_g, t_{g+1})$,
\[
\mathbf{W}_N(t, \hat{\boldsymbol{\theta}}) = \mathbf{W}_N(t, \boldsymbol{\theta}_0) - t_g\;\mathbf{W}_N(1, \boldsymbol{\theta}_0)+o_p(1)
   \longrightarrow_d \mathbf{J} \cdot t_g \cdot \delta \circ (\bar{\mathbf{h}}_g - \bar{\mathbf{h}})+ \{\mathbf{Z}(t)-t\cdot \mathbf{Z}(1)\}
\]

Thus under $H_1$,
\[
\mathbf{M}_N(t, \hat{\boldsymbol{\theta}}) = \hat{\mathbf{h}}^{-1/2}\mathbf{W}_N(t, \hat{\boldsymbol{\theta}})
   \longrightarrow_d \mathbf{J}^{1/2} \cdot t_g \cdot \boldsymbol{\delta} \circ (\bar{\mathbf{h}}_g - \bar{\mathbf{h}}) + \mathbf{W}^0(t)
\qquad t\in [t_g, t_{g+1})
\]
\end{proof}



\end{document}